# Wafer-scale all-epitaxial GeSn-on-insulator on Si(111) by molecular beam epitaxy

Krista R Khiangte[1], Jaswant S Rathore[1], J. Schmidt[3], H. J. Osten[3], A. Laha[2], S. Mahapatra[1]

[1]*Department of Physics, Indian Institute of Technology Bombay, Mumbai, INDIA*

[2]*Department of Electrical Engineering, Indian Institute of Technology Bombay, Mumbai, INDIA*

[3]*Institute of Electronic Materials and Devices, Leibniz Universität Hannover, Schneiderberg 32, D-30167 Hannover, GERMANY*

**Abstract**

In this letter, fabrication of all-epitaxial GeSn-on-insulator (GeSnOI) heterostructures is investigated, wherein both the GeSn epilayer and the $Gd_2O_3$ insulator are grown on Si(111) substrates by conventional molecular beam epitaxy. Analysis of the crystal and surface quality by high-resolution X-ray diffraction, cross-sectional transmission electron microscopy, and atomic force microscopy reveals the formation of a continuous and fully-relaxed single-crystalline GeSn epilayer (with a root-mean-square surface roughness of 3.5 nm), albeit GeSn epitaxy on $Gd_2O_3$ initiates in the Volmer-Weber growth mode. The defect structure of the GeSn epilayers is dominated by stacking faults and reflection microtwins, which are formed during the coalescence of the initially-formed islands. The concentration and mobility of holes, introduced by un-intentional p-type doping of the GeSn epilayers, were estimated to $8 \times 10^{16}$ $cm^{-3}$ and 176 $cm^{-2}V^{-1}s^{-1}$, respectively. In metal-semiconductor-metal Schottky diodes, fabricated with these GeSnOI heterostructures, the dark current was observed to be lower by a decade, when compared to similar diodes fabricated with GeSn/Ge/Si(001) heterostructures. The results presented here are thus promising for the development of these engineered substrates for (opto-)electronic applications.

**Introduction**

Owing to its tunable (direct) band gap and high carrier-mobilities, GeSn is rapidly emerging as a promising group-IV semiconductor for both photonic and electronic applications. Towards the long-standing goal of realizing optoelectronic components with group-IV alloys, fabrication of GeSn-based high speed photodetectors [1-3], light emitting diodes [4-6], and optically-pumped lasers have already been demonstrated [7] and significantly optimized [8]. In parallel, superior performance of GeSn-based *p*-channel metal-oxide-semiconductor-field-effect-transistors (*p*-MOSFETs) has been recently established [9-11], while optimization of *n*-MOSFETs [12, 13] and tunnelling-FETs [14-16] is being widely pursued. Naturally, the next step in GeSn-technology is the development of GeSn-on-insulator (GeSnOI) substrates, which is expected to combine the attractive material characteristics of the alloy with the "on-insulator" advantages [17]. However, the GeSnOI technology is still in its nascent stage, with only a few investigations reported so far [18-24].

Most of these efforts have focussed on formation of GeSn islands by liquid-phase epitaxy [18], or selected-area crystallization by pulsed laser annealing (PLA) of amorphous GeSn (*a*-GeSn) [19,20], or more complicated sandwich structures such as *a*-Ge/Sn/*a*-Ge heterostructures, grown on Si/SiO$_2$(001) [21] (and also capped with SiO$_2$), quartz [22], and glass substrates [23].

Fabrication of wafer-scale single-crystalline GeSnOI has been demonstrated only in Ref. [24], by direct wafer bonding (DWB). Apart from molecular beam epitaxy (MBE) of a Ge$_{0.96}$Sn$_{0.04}$/Si(001) epilayer, the technique involved several other processes, such as plasma-enhanced chemical vapour deposition of SiO$_2$, densification and post-bonding annealing, and chemical mechanical polishing. In this letter, we report conventional MBE growth of all-epitaxial GeSn/Gd$_2$O$_3$/Si(111) heterostructures on wafer scale, and analyse the crystal quality and microstructure of both the oxide and the alloy epilayers. We also demonstrate reduction of dark current in metal-semiconductor-metal (MSM) photodiodes fabricated with these heterostructures, which make them attractive templates for fabrication of group-IV based photonic components.

**Experimental**

In the first step of fabrication of the all-epitaxial GeSnOI heterostructures, ~ 15-nm-thick Gd$_2$O$_3$ epilayers were grown by MBE, on Boron-doped Si(111) substrates. The details of Gd$_2$O$_3$/Si(111) epitaxy, which has been thoroughly optimized over the last decade at Leibniz Universität Hannover, can be found in Ref. [25]. For the GeSn growth, the Gd$_2$O$_3$/Si(111) substrates were introduced in the ultrahigh vacuum (base pressure ~ $4.5 \times 10^{-10}$ mbar) growth chamber of the group-IV MBE system at IIT Bombay. Subsequently, the substrates were heated to 700 °C and annealed for ~ 5 mins. The GeSn epilayers were then grown at $T_G$ = 150 °C, at a growth rate of 1.4 nm min$^{-1}$. In-situ growth monitoring was done by RHEED (Staib Instruments Inc.), at a beam energy of 12 keV. High-resolution X-ray diffraction (HRXRD) scans and pole figures were recorded in a Rigaku Smartlab instrument, which is capable of performing scans in both out-of-plane and in-plane geometries. The diffractometer is equipped with a 9 kW rotating Cu anode, a parabolic mirror, a double-crystal Ge (220) monochromator, and several Soller slits. While, wide ω-2θ scans were performed in the high resolution mode, wherein the two-bounce monochromator was used on the incident side, the θ-2θ and the pole figure measurements were carried out in a low resolution mode, with a 5.0 degree Soller slit replacing the monochromator.

For the pole figure measurements, the sample tilt (χ) was varied from 0 ° (in which the scattering vector is parallel to the surface normal) to 90 ° (in which the scattering vector lies in the same plane) while the azimuthal angle (ɸ) was varied by sample rotation about the surface normal. Additionally, θ-2θ scans have also been recorded for specific (χ,ɸ) spots of the pole figure. The microstructure, stacking configurations, and defect structure were also analysed by recording cross-sectional transmission electron microscopy (XTEM) images along the $[1\bar{1}0]$ direction, using a JEOL 200 microscope, operating at voltages up to 200 kV. For the fabrication of metal-semiconductor-metal (MSM) back-to-back Schottky diodes, a 350-nm-thick $SiO_2$ passivation layer was grown at 200 ºC, atop the epi-Ge layer, by inductively-coupled-plasma-assisted chemical-vapour-deposition (ICPCVD). Subsequently, inter-digitated electrodes (IDE) were patterned by single level optical lithography and deposition of Ni (20 nm)/Au (50 nm) contacts by electron beam evaporation (see inset of Figure 4). The current-voltage characteristics of the MSM structures were measured at room temperature with an Agilent B1500A Semiconductor Device Parameter Analyser.

**Results and discussion**

Figure 1 shows the evolution of the reflection high energy electron diffraction (RHEED) pattern during the growth of the GeSn epilayers. The streaky RHEED pattern of Figure 1(a), recorded immediately before the commencement of GeSn growth, indicates that the high-temperature annealing resulted in the formation of a clean and smooth epi-$Gd_2O_3$ starting surface. As the GeSn growth began, the diffuse background intensity increased, thus reducing the visibility of the streaks (Figure 1(b)). This evolution is most likely related to the formation of an amorphous Ge/Sn-oxide layer [26]. The pattern changed substantially after about 1.5 min of growth, when a spotty pattern, as shown in Figure 1(c), emerged. The spotty pattern may be attributed to the onset of GeSn island formation in the Volmer-Weber (V-W) growth mode. Figure 1(d), which shows a close up image of the white box in Fig. 1(c), reveals two super-imposing spotty patterns, wherein one pattern (dotted lines) is the mirror-image of the other one (solid lines), reflected about the (00L) rod. This indicates the presence of GeSn islands with both type-A and type-B stacking of the (111) planes [27, 28]. Here type-A refers to the stacking sequence of the (111) planes in the Si(111) substrate (for example, ABCABC…), while type-B refers to stacking of the (111) planes, rotated azimuthally (i.e. about the [111] axis) by 180 ° (i.e. BACBAC…). Additional spots at fractional L positions suggest the presence of reflection twins, which will be discussed in detail later. The superimposing spotty pattern persisted for up to 3 hours of GeSn growth (corresponding to GeSn deposition equivalent to a layer thickness of ~250 nm). Beyond this duration of growth, one of the two superimposing spotty patterns completely disappeared (Figure 1(e)).

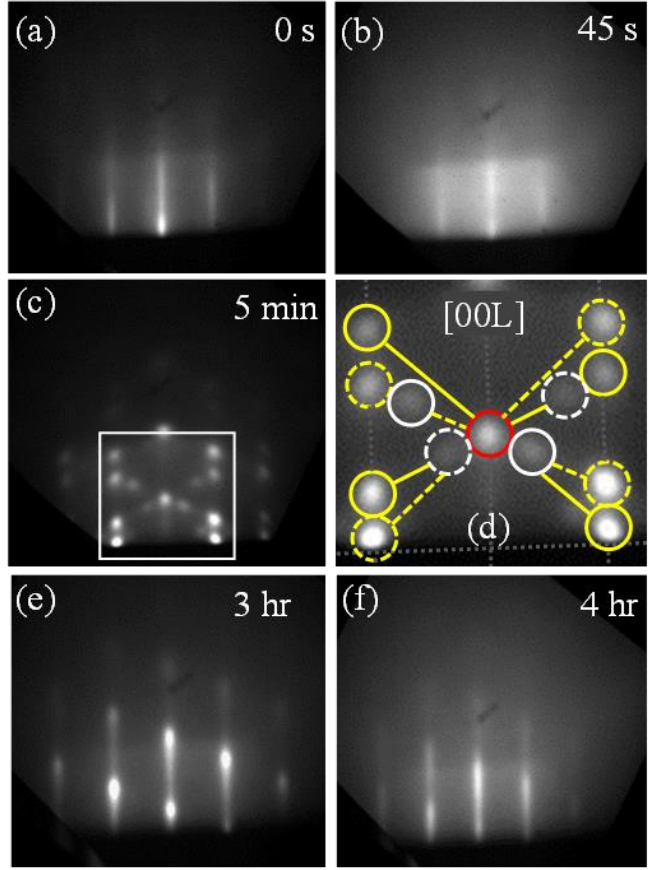

*Figure 1: RHEED images recorded at different stages of GeSn growth: (a) Image of the $Gd_2O_3$(111) surface just before the start of GeSn growth. Images recorded after (b) 45 s, (c) 5 min, (e) 3 hours and (f) 4 hours of GeSn growth. (d) A close-up image of the area demarcated by the white box in (c). The spots within yellow dotted circles are mirror images (about the (00L) rod) of the spots within the yellow solid circles, suggesting the presence of rotational twins. The spots encircled in white are due to reflection microtwins.*

This indicates that in the subsequent growth, one of the two domains grew preferentially. Finally, a streaky pattern emerged after 4 hours, suggesting the recovery of two-dimensional (2D) growth. It is thus evident that while GeSn growth on $Gd_2O_3$ begins in the V-W growth mode, a 2D surface is recovered for sufficiently thick layers. The evolution of the RHEED pattern described above is very similar to that reported earlier for Ge epitaxy on epi-$Pr_2O_3$/Si(111) [26] and epi-$Gd_2O_3$/Si(111) [28] substrates.

The single-crystalline nature of both the $Gd_2O_3$ and the GeSn epilayers is established by the out-of-plane wide $\omega$-$2\theta$ diffractogram of Figure 2(a), which was recorded from a sample with a ~ 670-nm-thick GeSn layer. The $2^{nd}$, $4^{th}$ and $6^{th}$ order of the (111) reflections of $Gd_2O_3$ are observed to nearly coincide with the $1^{st}$, $2^{nd}$ and $3^{rd}$ order of the (111) reflections of Si, since the bulk lattice constant of the former (10.812 Å) is only slightly smaller than twice the bulk lattice constant of the latter (5.431Å) [29].

Additionally, reflections from only the (111) and the (333) planes of the GeSn epilayer are visible over the entire range of 2θ values (22° to 100°), proving that the GeSn layer is fully single-crystalline. The right inset shows the ω rocking-curve diffractogram for the GeSn (111) peak, revealing a full-width-at-half-maximum (FWHM) value of 0.57°. This value is twice that of the FWHM measured for Ge epilayers grown on the same substrates earlier [30]. To determine the lattice constant(s) and the strain state of the GeSn epilayer, skew symmetric θ-2θ scans were recorded for the (224), (220) and ($0\bar{2}6$) reflections. The left inset of Figure 2(a) shows the high-resolution θ-2θ diffractogram corresponding to the (220) reflections. The calculated values of the in-plane and out of-plane lattice constants reveal that the GeSn epilayer is completely relaxed. The Sn composition estimated using Vegard's law [31] is 7.0%.

An atomic-force-microscope (AFM) image of the same sample is shown in Figure 2(b). The surface appears to be predominantly flat, with a root mean square (RMS) roughness of 3.5 nm. In contrast, the surface of the thinner GeSn epilayer (210 nm) of the same composition, shown in Figure 2(c), reveals the presence of large islands. This directly supports the observations made earlier in the context of RHEED analysis, regarding the growth evolution of GeSn on $Gd_2O_3$/Si(111). However, the measured RMS surface roughness (of the thicker layer) is nearly double that of GeSnOI(001) substrates fabricated by DWB (~ 1.9 nm), with much lower thickness (100 nm) and Sn-content (4.1 %) of the GeSn epilayers [24].

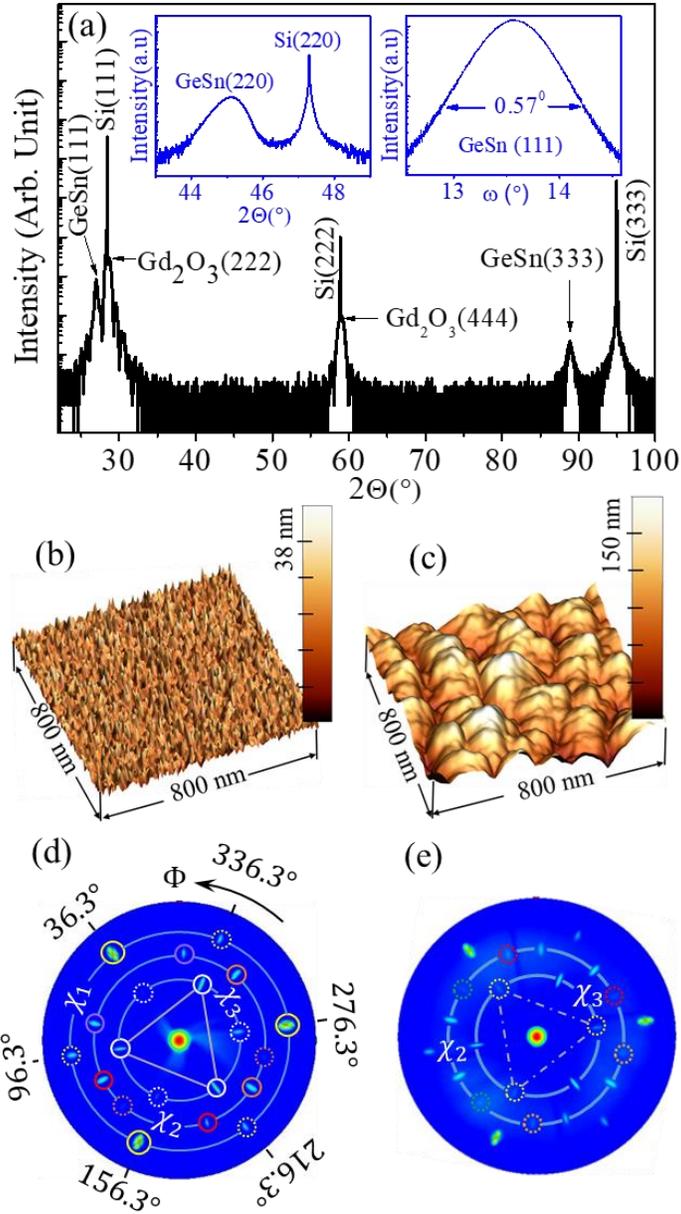

Figure 2: (a) wide ω-2θ diffractogram recorded from the 670-nm-thick GeSn/Gd$_2$O$_3$/Si(111) heterostructure. The left inset shows a high-resolution θ-2θ diffractogram recorded for the (220) reflection whereas the right inset shows the ω-rocking curve diffractogram corresponding to the GeSn (111) reflection. AFM image of the surface of (b) a 670-nm-thick and (c) a 210-nm-thick GeSn/Gd$_2$O$_3$/Si(111) sample. XRD - pole figure recorded at the Bragg angle corresponding to the GeSn (111) reflection for (d) the thicker and (e) the thinner GeSn epilayer. In (d), the equi-tilt (χ) contours are shown schematically by the concentric circles. The centre of the pole figure corresponds to χ = 0, while the circumference corresponds to χ = 90°. A few azimuthal angles (ϕ) are also marked. (e) XRD - pole figure measurement at the Bragg angle corresponding to the GeSn(111) reflection for sample with GeSn layer thickness of 210 nm.

Figure 2(d) shows the XRD pole figure recorded for the thicker GeSn epilayer at the Bragg angle ($2\theta = 27.0°$) of the GeSn (111) reflection, which provides details of the epitaxial relationship between the epilayers (and the substrate), and the defect structure on a global scale. The centre of the pole figure corresponds to no sample tilt ($\chi = 0$), which implies that the reflection here stems from the GeSn (111) planes. The triad of reflections at $\chi_1 = 71.5°$, encircled in solid yellow, correspond to the three $\{\bar{1}11\}$ crystal planes. Together with the (111) spot at the centre, these three spots (which are separated azimuthally by $\Delta\Phi = 120°$), reflect the 3-fold symmetry of the GeSn crystal, about the [111] axis. However, another triad of reflections, marked by yellow dotted circles, is also visible at the same tilt ($\chi_1 = 71.5°$). It is easy to notice that the latter triad can be obtained by a 180° azimuthal rotation of the former. This indicates the co-existence of both type-A (solid yellow) and type-B (dotted yellow) stacked domains in the GeSn epilayer [32]. By calculating the ratio of the peak- areas in $\theta$-$2\theta$ diffractograms, corresponding to the spots at $\chi_1 = 71.5°$, $\phi = 36.2°$ and $\chi_1 = 71.5°$, $\phi = 96.2°$, the volume fraction of the GeSn in the type-B domains is estimated to be only ~ 1.2%.

The triad of reflections at $\chi_3 = 38.9°$, marked by solid white circles correspond to reflection microtwins of the (111) planes in type-A domains, which are reflected about the three $\{\bar{1}11\}$ planes. The pairs of reflections at $\chi_2 = 54.5°$, (marked by red, purple, and orange circles) represent the triads whose third apices lie outside the detection range ($\chi = 109.5°$) [32]. These triads are symmetric equivalents of the triad at $\chi_3 = 38.9°$, and hence correspond to the same reflection microtwins (i.e. within the type-A domains).

Also visible in the same image, although very faintly, is a triad of spots at $\chi_3 = 38.9°$, (marked by dotted white circles). Once again, this triad is rotated by 180° with respect to the other triad at the same $\chi$ (solid white circles), which proves that the former triad corresponds to reflection microtwins in type-B domains. While these spots are barely visible in the pole-figure of the thicker sample (Fig. 2 (d)), they are significantly more pronounced in Figure 2(e), which shows a similar pole-figure recorded for the thinner sample. Additionally, Fig. 2(e) shows three pairs of reflections (marked by red, green, and orange dotted circles) at $\chi_2 = 54.5°$, which were not recorded for the thicker sample. As explained in the context of Fig. 2(d), these reflections also correspond to reflection microtwins, but in type-B domains. The fact that the Bragg spots due to type-B-oriented reflection twins are (more prominently) seen in the XRD-pole figure of the 210-nm-thick-GeSn-epilayers, strongly suggests that the type-B domains are restricted to within a certain thickness of the epilayer close to the GeSn-$Gd_2O_3$ interface, beyond which only type-A GeSn grows. The pole figures presented above are similar to those reported for Ge/epi-$Pr_2O_3$/Si(111) epitaxy in Ref. [32], where further details may be found regarding the method of analysis.

The XTEM images shown in Figure 3 provide further insight to the crystallinity of the GeSn epilayers and the microstructure of the defects. Figures 3(a) (Figure 3(b)) show the $[1\bar{1}1]$ cross-sections over relatively large areas, for the sample with the 670-nm-thick (210-nm-thick) GeSn epilayer. As suggested by the RHEED evolution, the thinner layer is characterized by a 3D surface, consisting of well-facetted islands. The thicker layer on the other hand is closed, and characterized by a relatively smooth 2D surface. However, in both samples, stacking faults (SFs)/twins are observed to propagate through the entire thickness of the GeSn epilayers. In thicker GeSn epilayers (~ 1 μm), the propagation of stacking faults is expected to be arrested [32], which would lead to higher crystalline quality close to the sample surface.

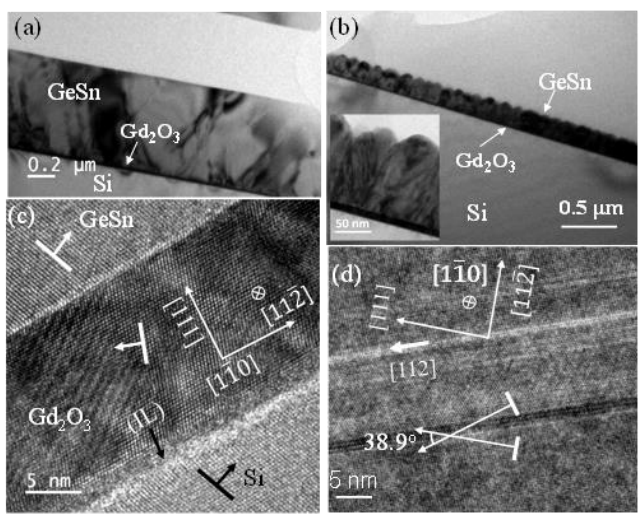

*Figure 3: XTEM image of a GeSn/Gd2O3 /Si(111) sample with (a) a 670-nm-thick and (b) a 210-nm-thick GeSn epilayer. The inset of (b) shows the facetted islands at the surface of the thinner epi-GeSn layer. (c) A high resolution image of the interface region, showing the stacking orientation of the $Gd_2O_3$ and the GeSn epilayers, relative to that of the Si(111) substrate. The arrows (in white) indicate the normal to the ($11\bar{1}$) planes in each region. (d) A high-resolution image of a reflection microtwin lamella in 670-nm-thick GeSn epilayer.*

A high-resolution image of the Ge/epi-$Gd_2O_3$/Si interface is shown in Figure 3(c). The normal to the ($11\bar{1}$) planes, indicated for each of the Si, $Gd_2O_3$, and GeSn regions by arrows, reveal that the epi-GeSn and epi-$Gd_2O_3$ layers have a type-A and type-B stacking order, respectively, w.r.t. to the Si substrate [33]. The type-B stacking of epi-$Ln_2O_3$ layers is well-established from previous studies [32]. Also, the A/B/A stacking of the entire heterostructure has been reported earlier for Ge epitaxy on $Gd_2O_3$/Si(111) [27] and $Pr_2O_3$/Si(111)

[32]. In the same image, an amorphous interfacial layer (IL) is visible at the $Gd_2O_3$-Si interface, which is possibly formed during the high temperature annealing of the epi-$Gd_2O_3$/Si(111) surface [34]. Figure 3(d) depicts one of the reflection-microtwin regions. Within the lamella at the middle of the image, the (111) planes are observed to be tilted with respect to the global (111) planes by an angle of 38.9°. This suggests that the microtwin lamellae are formed by reflection of un-faulted (111) planes about the $(11\bar{1})$ twin plane [32], as previously argued in the pole figure analysis.

Hall measurement carried out in the van der Pauw geometry at room temperature revealed that the GeSn/$Gd_2O_3$/Si(111) heterostructures were p-doped. This unintentional doping is possibly contributed by shallow acceptor levels, formed due to the presence of stacking faults/threading dislocations in the grown GeSn epilayer [35]. From the plot of the measured Hall voltage versus the applied magnetic field (right inset of Figure 4), the hole concentration was measured to be $p = 8 \times 10^{16}$ cm$^{-3}$. The carrier mobility was then estimated (from the longitudinal voltage) to be $\mu = 176$ cm$^{-2}$V$^{-1}$s$^{-1}$. It is worth noting that the mobility value is only slightly smaller than that reported ($\mu = 208$ cm$^{-2}$V$^{-1}$s$^{-1}$) for the first $p$-FINFETs fabricated with DWB-GeSnOI (001) substrates [36].

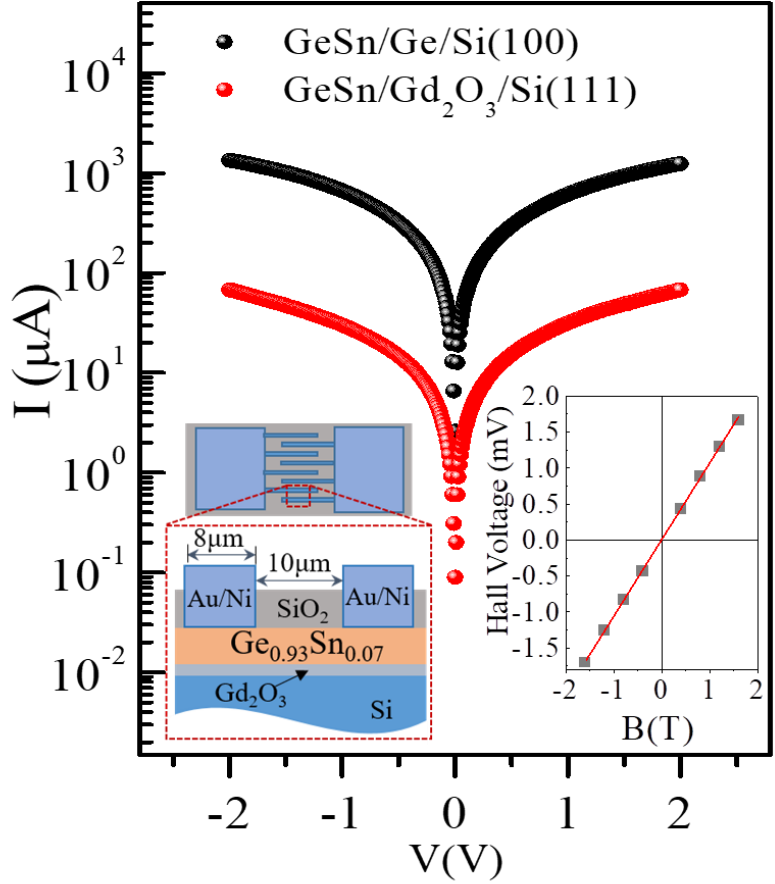

*Figure 4: I-V characteristics of MSM photodiodes fabricated with a $Ge_{0.93}Sn_{0.07}$ (670 nm)/$Gd_2O_3$ (10 nm)/Si(111) heterostructure (black) compared to that of MSM photodiodes fabricated with a $Ge_{0.93}Sn_{0.07}$ (650 nm)/Ge (300 nm)/Si(001) heterostructure (red). The left inset shows a schematic top-view (top) and cross-section (bottom) of the interdigitated electrodes of the MSM photodiodes. The right inset shows the variation of Hall voltage with the applied magnetic field (B).*

The main panel of Fig. 4 compares the measured dark current of MSM back-to-back Schottky diodes fabricated with a $Ge_{0.93}Sn_{0.07}$ (670 nm)/$Gd_2O_3$/Si(111) heterostructure, to that of similar diodes fabricated with a $Ge_{0.93}Sn_{0.07}$ (650 nm)/Ge (300 nm)/Si(001) heterostructure, also grown in the same MBE chamber. The details of the GeSn/Ge/Si(001) growth can be found in ref [37]. The data reveals a clear reduction in the dark current by an order of magnitude in the case of the former. This may be attributed to the isolation of the Si substrate due to the introduction of epi-$Gd_2O_3$ insulating layer.

**Conclusions**

In conclusion, wafer-scale fabrication of all-epitaxial GeSnOI(111) engineered substrates, by MBE growth of $Gd_2O_3$ and $Ge_{0.93}Sn_{0.7}$ epilayers on Si(111) substrates has been demonstrated in this work. The crystalline quality, surface morphology, defect structure, and electrical properties of the heterostructures have been systematically investigated by HRXRD, pole figure analysis, XTEM and AFM, while the carrier concentration and mobility have been estimated by Hall measurement. Reduction of dark current in MSM photodiodes fabricated with the GeSnOI heterostructures has also been observed. This finding is promising for future development of these GeSnOI templates for photonic applications.

**Acknowledgement**

The research was funded by the Science and Engineering Research Board, Department of Science and Technology (DST), Government of India. KRK, JR, AL, and SM acknowledge support from the Centre of Excellence in Nanoeletronics (CEN) and Industrial Research and Consulting Centre (IRCC), Indian Institute of Technology Bombay.